\documentclass[12pt]{article}
\usepackage{amsmath,amssymb}
\usepackage{epsfig}
\newcommand{\be}{\begin{equation}}
\newcommand{\ee}{\end{equation}}
\newcommand{\bea}{\begin{array}}
\newcommand{\ea}{\end{array}}
\newcommand{\beqa}{\begin{eqnarray}}
\newcommand{\eeqa}{\end{eqnarray}}

\begin{document}


\begin{flushright}   SU-4252-807 \\ 
                     SINP/TNP/05-02 \\
                     DIAS-STP-05-03 \\ 
\end{flushright}         
\begin{flushright}   
                     September 2005 \\
\end{flushright} 
\begin{center}
\vskip 3em
{\LARGE Waves on Noncommutative Spacetimes}
\vskip 3em
{\large A. P. Balachandran$^a$, Kumar S. Gupta$^b$\footnote{Regular
Associate, Abdus Salam ICTP, Trieste, Italy} and 
S. K\"{u}rk\c{c}\"{u}o\v{g}lu$^{c}$\footnote{E-mails: bal@phy.
syr.edu, gupta@theory.saha.ernet.in, seckin@stp.dias.ie}} \\[3em]

\em{$^a$ Department of Physics, Syracuse University, Syracuse, NY 13244-1130, USA. \\
\vskip 1em
$^b$ Theory Division, Saha Institute of Nuclear Physics, 1/AF Bidhannagar,\\ 
Kolkata 700064, India.
\vskip 1em
$^c$ Dublin Institute for Advanced Studies, School of Theoretical 
Physics,\\
10 Burlington Road, Dublin 4, Ireland.}

\end{center}

\vskip 2em

\section{Introduction}
 
Studies on the formulation of physical theories on the Moyal plane were initiated in recent times by Doplicher et al. 
\cite{doplicher}. Interest in such algebras was also stimulated by the work of string theorists who encountered 
them in a certain decoupling limit \cite{Seiberg}.

The $d$-dimensional Groenewold-Moyal spacetime is an algebra ${\cal A}_\theta({\mathbb R}^d$) generated by 
elements ${\hat x}_\mu$ ($\mu \in \lbrack 0,1,2, \cdots, d-1 \rbrack$) with the commutation relation
\be
\lbrack {\hat x}_\mu \,, {\hat x}_\nu \rbrack = i \theta_{\mu \nu} {\bf 1}\,,
\ee

$\theta_{\mu \nu}$ being real constants antisymmetric in its indices. In the limit $\theta_{\mu \nu} = 0$, ${\hat x}_0$ 
and ${\hat x}_i$ are time- and space- coordinate functions. If $x = (x_0 \,, \vec{x})$ is a point of ${\mathbb R}^d$ when
 $\theta_{\mu \nu} = 0$, then
\be
{\hat x}_0(x) = x_0 \,, \quad {\hat x}_i(x) = x_i \,.
\ee
Thus ${\hat x}_0$, ${\hat x}_i$ are operators in ${\cal A}_\theta({\mathbb R}^d$) which become time and space coordinate 
functions when $\theta_{\mu \nu} = 0$. There is an extensive literature on the formulation of quantum field theories 
($qft's$) on ${\cal A}_\theta({\mathbb R}^d)$ and on their phenomenology \cite{IJMPArev}. The focus of much of 
this work is on space-space noncommutativity ($\theta_{ij} \neq 0, ~ i,j \in [1,2,...d-1]$).
But it is time-space noncommutativity ($\theta_{0i} \neq 0$) with its implications for causality 
and foundations of quantum theory which leads to strikingly new physics.  The formulation of unitary $qft's$ when
$\theta_{0i} \neq 0 $ is nontrivial and was carefully done already by Doplicher et al. \cite{doplicher}.

In  recent papers \cite{uqp, Bal2}, Balachandran et al. formulated unitary quantum mechanics when $\theta_{0i} \neq 0$ 
basing themselves on the ideas of Doplicher et al. Consequences of spacetime noncommutativity in the quantum mechanics of
atoms and molecules have been explored by Balachandran and Pinzul \cite{BalSasha}.

Previous work on ${\cal A}_\theta({\mathbb R}^d)$ was focused on the formulation of quantum theory. Effects of 
noncommutativity on classical waves and particles have largely remained untreated in particular when $\theta_{ij} 
\neq 0$ (see however \cite{Nair}). In this paper we discuss ``classical'' waves on ${\cal A}_\theta({\mathbb R}^d$), 
assuming time-space noncommutativity ($\theta_{ij} = 0 \,, \theta_{0i} \neq 0$).

The approach adopted in Doplicher et al. \cite{doplicher}  and subsequently in 
Balachandran et al. \cite{uqp} to study space-time noncommutativity is different 
from the string theory motivated studies in the literature due to Gomis and Mehen  
\cite{Gomis} and other authors \cite{Others}, which found that field theories on noncommutative 
spacetime are perturbatively nonunitary. As explained in detail in Balachandran et al. \cite{uqp},
in the former approach, the amount $\tau$ of time translation is not ``coordinate time'', 
the eigenvalue of $\hat{x}_{0}$. 
For $\theta=0$, these two could be identified, while for 
$\theta\ne0$, $\hat{x}_{0}$ is an operator not commuting with $\hat{x}_{1}$, and cannot be 
interchanged with $\tau$. 
The separation of eigenvalues of $\hat{x}_{0}$ from the amount of time
translation is the central reason for the unitarity of the theories as formulated in 
Doplicher et al. and Balachandran et al \cite{doplicher, uqp}. 
This is analogous to the situation in quantum 
mechanics, where if $\hat{p}$ is the momentum operator, spatial translation by amount $\xi$ 
implemented by $\exp(i\xi\hat{p})$ is not the eigenvalue of the position operator $\hat{x}$.

In the algebraic approach (which is mandatory if $\theta_{\mu \nu} \neq 0$), waves are elements of the spacetime algebra.
That is the case also for the commutative space-time $C^{0}({\mathbb R}^d)$. The act of observation, such as measurement
of mean intensity over the time interval $T$, is represented by a state on this algebra. In the following section, we 
describe this approach, valid equally for commutative and noncommutative algebras. Subsequently it is applied to
interference for a double slit experiment for the algebra ${\cal A}_\theta({\mathbb R}^d)$. In cases where a double image
of a star is formed by a cosmic string, it causes interference as well which is affected by $\theta_{0i}$. This 
phenomenon is examined in the final section.

Novel phenomena are observed in interference when $\theta_{0i} \neq 0$. For instance if the time of observation $T$ 
is too small, then, as indicated before, one sees constant intensity and no interference on the screen. Interference 
returns for larger times, but it is shifted and deformed as a function of $\frac{\theta w}{T}$ 
($\theta = \sqrt{\Sigma_i \theta_{0i}^2} \,, w$ : frequency of the wave). The familiar interference pattern is recovered 
only when $\frac{\theta w}{T} \rightarrow 0$.

\section{Classical Waves and Particles on Algebras}

The algebra ${\cal A}_\theta({\mathbb R}^d)$ has generators ${\hat x}_\mu$ with relations 
$\lbrack {\hat x}_\mu \,, {\hat x}_\nu 
\rbrack = i \theta_{\mu \nu}$, $ \theta_{\mu \nu} = - \theta_{\nu \mu}$ being real constants. We assume that 
$\theta_{ij} = 0$, and orient $\vec{\theta}_{0} = ( \theta_{01}\,, \cdots \theta_{0 d-1})$ in some direction $\vec{n}$.

Thus, for us, 
\be 
\lbrack {\hat x}_i \,, {\hat x}_j \rbrack = 0 \,, \quad
\lbrack {\hat x}_0 \,, {\hat x}_i \rbrack = i \theta n_i \,, 
\quad \theta \in {\mathbb R}, ~~\vec{n} \cdot \vec{n} = 1 \,.
\label{eq:first}
\ee

We can  set $\theta \geq 0$. This does not entail loss of generality since $\theta$ flips in sign when 
${\hat x}_i \longrightarrow -{\hat x}_i$. So $\vec{\theta}_0 = \theta \vec{n}$, $\theta \geq 0$. 

For $\theta = 0$, ${\cal A}_\theta({\mathbb R}^d)$ is the algebra $C^0({\mathbb R}^d)$ of functions on ${\mathbb R}^d$.
We first outline the algebraic approach for $\theta = 0$. It generalizes easily to $\theta \neq 0$.

\vskip 0.3cm

{\it i.~ Classical Theory on Commutative Algebra:} \\

Let us first examine waves. They are fields ${\hat \psi}$ on spacetime so that 
${\hat \psi} \in {\cal A}_0({\mathbb R}^d)$. It is 
enough to consider scalar waves. Then ${\hat \psi}(x)$ for $x= (x_0 \,, \vec{x}) \in {\cal A}_0({\mathbb R}^d)$ is the 
amplitude of the wave ${\hat \psi}(x)$ at $x$. It is the solution of a wave equation such as
\be
\left ( \partial_0^2 - \sum_1^{d-1} \partial_i^2 \right ) {\hat \psi}(x) = 0 \,. 
\ee       

The algebra ${\cal A}_0({\mathbb R}^d)$ contains not just ${\hat \psi}$, but functions of ${\hat \psi}$ as well. It
is reasonable to assume that a general element ${\hat \alpha} \in {\cal A}_0({\mathbb R}^d)$ has the Fourier 
representation
\be
{\hat \alpha} = \int d^d k \, {\tilde \alpha}(k) \, e^{i k_0 {\hat
    x}_0} e^{i \vec{k} \cdot {\vec {\hat {x}}}} \,,
\ee
where ${\hat x}_\mu$ are the coordinate functions: ${\hat x}_\mu (x) = x_\mu$.

We can measure many attributes of a wave. For example, we can measure its mean intensity over the time interval
$\lbrack x_0 -\frac{T}{2} \,, x_0 + \frac{T}{2} \rbrack$. It is 
\be
I = \frac{1}{T} \int_{x_0 - \frac{T}{2}}^{x_0+\frac{T}{2}} dx_0 \, |{\hat \psi}(x_0 \,, \vec{x})|^2 \,.
\label{eq:ints1}
\ee
We want to interpret this measurement as the application of a state on a particular element of the algebra since 
states are defined also for noncommutative algebras.

A state $\omega$ on a $*$-algebra ${\cal A}$ with unity ${\bf 1}$ is a linear map \cite{Madore},
\be
\omega : {\hat \alpha} \in {\cal A} \longrightarrow {\mathbb C}
\ee
which is positive
\be
\omega ({\hat \alpha}^* {\hat \alpha}) \geq 0
\ee     
and normalized :
\be
\omega({\bf 1}) = 1 \,.
\ee
Thus states define probabilities and $ \omega ({\hat \alpha})$ is the mean value of ${\hat \alpha}$.

Coming back to (\ref{eq:ints1}), for intensity, we associate the observable {\bf ${\hat I} 
\in {\cal A}_0({\mathbb R}^d)$} where
\be 
{\hat I} = |{\hat \psi}|^2 \,.
\ee  

Measurement of the mean value of ${\hat \alpha} \in { {\cal A}_0}({\mathbb R}^d)$ at $\vec{x}$ in the time-interval  
$\lbrack x_0-\frac{T}{2} \,, x_0+\frac{T}{2} \rbrack$ is represented by the state $\omega$ where
\be 
\omega({\hat \alpha}) = \frac{1}{T} \int_{x_0- \frac{T}{2}}^{x_0+\frac{T}{2}} dx_0 \, {\hat \alpha}(x_0\,, {\hat x}) \,.
\label{eq:conts}
\ee
So $\omega$ depends on $x_0, T$ and $\vec{x}$. Then
\be
I = \omega({\hat I}) \,.
\ee
Thus to an observable, we assign an element ${\hat \alpha} \in {\bf {\cal A}_0}({\mathbb R}^d)$ and to a measurement, 
a state $\omega$ on ${\bf {\cal A}_0}({\mathbb R}^d)$. The result of this measurement of ${\hat \alpha}$ is 
$\omega({\hat \alpha})$.

A classical particle too can be described by a similar formalism. Instead of working with ${\mathbb R}^d$, it is best
to include momenta also and work with ${\mathbb R}^{2d-1}$. A point of ${\mathbb R}^{2d-1}$ is $(x_0, \vec{x}, \vec{p})$ 
where $\vec{p}$ denotes momenta. The algebra is then ${\cal A}_0({\mathbb R}^{2d-1})$. If ${\hat \alpha} \in 
{\cal A}_0({\mathbb R}^{2d-1})$,  ${\hat \alpha} (x_0, \vec{x}, \vec{p})$ is the value of the observable ${\hat \alpha}$ 
at time $x_0$ for a particle with position $\vec{x}$ and momentum $\vec{p}$. For energy ${\widehat E} \in {\cal A}_0
({\mathbb R}^{2d-1})$ in a possibly time-dependent potential, we can have ${\widehat E}(x_0 \,, \vec{x}, \vec{p}) = 
\frac{\vec{p}^2}{2m} + V(x_0 \,, \vec{x})$.

States too can be introduced. For example, define $\omega$ by
\be
\omega({\hat \alpha}) = \frac{1}{T} \int_{x_0 -\frac{T}{2}}^{x_0+\frac{T}{2}} d x_0 \, {\hat \alpha}(x_0 \,, \vec{x} \,, 
\vec{p}) \,.
\ee
So $\omega$ depends on $x_0, T, \vec{x}$ and $\vec{p}$.  $\omega({\widehat E})$ is the mean energy in the time interval
$\lbrack x_0- \frac{T}{2}\,, x_0 + \frac{T}{2} \rbrack$ for a particle at $\vec{x}$ with momentum $\vec{p}$. 

We, however, will not pursue point-particle theory any further. 

\vskip 0.3 cm

{\it ii. ~ Classical Waves on Noncommutative Algebra ${\cal A}_\theta({\mathbb R}^d)$} \\

As remarked already, states can be defined also on ${\cal A}_\theta({\mathbb R}^d)$. Thus to carry the discussion
forward, we must identify waves in ${\cal A}_\theta({\mathbb R}^d)$ say by wave equations, associate observables
to waves and define suitable states. We will do so in the context of interference and diffraction for $d \leq 4$
in what follows. But we must emphasize one
strikingly new feature of $\theta \neq 0$. Then since ${\hat x}_0$ and ${\hat x}_i$ do not commute, by the uncertainty 
principle, we can not simultaneously localize time in an interval $T$ and sharply localise spatial coordinates. So a 
state like $\omega$ in (\ref{eq:conts}) with exactly the same features does not exist for $\theta \neq 0$. We can at 
best approximate it.

We consider free massless scalar fields ${\hat \psi} \in {\cal A}_\theta({\mathbb R}^d)$ for $d= 2,3,4$. Such massless 
scalar fields obey the standard wave equation
\be
(\partial_0^2 - \vec{\nabla}^2) {\hat \psi} = 0 
\ee
for $\theta =0$. We must find its analogue for $\theta \neq 0$. 

For simplicity, we choose $\vec{n} = (1, 0, 0)$, if necessary by applying a spatial rotation on ${\hat x}_i$.

Let
\begin{gather}
{\widehat P}_0 = - \frac{1}{\theta} ad \, {\hat x}_1 \nonumber \\
ad \, {\hat \alpha} ~ {\hat \beta} := \lbrack {\hat \alpha} \,, {\hat \beta} \rbrack \,.
\end{gather}
Then 
\be
{\widehat P}_0 {\hat x}_0 = i \,, \quad  {\widehat P}_0 {\hat x}_i = 0,~ i \geq 1 \,.
\ee
So ${\widehat P}_0$ substitutes for $i \frac{\partial}{\partial x_0}$ and we can identify $\partial_0$ with $-i
{\widehat P}_0$:
\be
\partial_0 \longrightarrow -i {\widehat P}_0 \,.
\ee
Similarly
\be
\partial_1 \longrightarrow  i {\widehat P}_1 =  - \frac{i}{\theta} ad \, {\hat x}_0
\ee
while 
\be 
\partial_a \longrightarrow \partial_a : = i {\widehat P}_a \,, \quad a = 2, 3, \,
\ee
$\partial_a$ in (19) being the conventional differentiations. So the noncommutative elementary massless wave equation 
is 
\be
({\widehat P}_0^2 - {\widehat P}_1^2 - {\widehat P}_a^2) {\hat \psi} = 0 \,.
\label{eq:waveq1}
\ee
It has plane wave solutions
\be
{\hat \psi}_{\vec{k}} = e^{i k_i {\hat x}_i} e^{-i w {\hat x}_0} \,,
\label{eq:wavsol1}
\ee
with a standard dispersion relation:
\be
w^2 - \vec{k}^2 =0 \,. 
\ee     
The general solution is a superposition of plane waves.

We note that for electromagnetic (EM) waves, (\ref{eq:waveq1}) receives corrections
in powers of $\theta$, since in noncommutative spacetimes, the EM
Lagrangian gives nonlinear equations of motion. Interestingly enough monochromatic plane waves of the form 
(\ref{eq:wavsol1}) are solutions to the nonlinear equations of motion to every order in $\theta$ \cite{berrino}. 
But their superposition is not. Nevertheless, as the inclusion of this effect will give 
only higher order corrections in $\theta$ to our results, they are not treated in this paper.  
Similarly the use of ``covariant coordinates" (cf. \cite{zahn} and references therein) 
will not affect leading order results in $\theta$ and hence we work with the standard noncommutative coordinates.  

\vskip 0.3cm

{\it a) $d=2$}: 

\vskip 0.2cm

The problem we examine is the interference of two plane waves with the same frequency. It can be generalized, but several
essential points are well illustrated by this example. Thus we consider
\begin{gather}
{\hat \psi} = {\hat \psi}_k +{\hat \psi}_{-k} \,, \nonumber \\
{\hat \psi}_{\pm k} = e^{\pm i k {\hat x}_1} e^{-i k {\hat x}_0} \,, \quad k >0 \,.
\label{eq:planewave}
\end{gather}

We see that the intensity 
\be
{\hat I} = |{\hat \psi}|^2
\ee
is an operator in ${\cal A}_\theta ({\mathbb R}^2) \,.$

It is not possible to achieve a state with a sharp localization in position and time. Instead, we look for a state 
$\omega$ with a reasonable spatial localization around a point $x_1$, it will be rather delocalized in time. We define 
$\omega \equiv \omega_\gamma$ in terms of a density matrix according to 
\be
\omega_\gamma ( {\hat \alpha}) = \frac{Tr {\hat \gamma} {\hat \alpha}}{Tr {\hat \gamma} ({\bf 1})} \,,
\label{eq:densitymat}
\ee  
where
\be
{\hat \gamma} = {\hat \psi}_T({\hat x_0}- x_0) {\hat \delta} ( {\hat x}_1 - x_1) \, {\hat \psi}_T({\hat x_0}- x_0) \,.
\label{eq:gamma}
\ee
Here by $ {\hat \delta}$ and ${\hat \psi}_T$ we mean the following operators:
\beqa
{\hat \delta}({\hat x}_1 - x_1) &=& 
\frac{1}{2 \pi} \int d k \, e^{i k ({\hat x_1} - x_1)}  \,, \label{eq:deltaf} \\
{\hat \psi}_T ({\hat x_0}- x_0) &=& 
\int^{\frac{T}{2}}_{-\frac{T}{2}} d
\lambda~ {\hat \delta}({\hat x}_0 - x_0 - \lambda)\,. \label{eq:distpsi}
\eeqa

Let $|x_1^\prime \rangle$ be the eigenstate of ${\hat x}_1$ for the eigenvalue $x_1^\prime$:
\be 
{\hat x}_1 |x_1^\prime \rangle = x_1^\prime |x_1^\prime \rangle \,.
\ee
Then
\be
{\hat \delta} ( {\hat x}_1 - x_1) |x_1^\prime \rangle =  \delta ( x_1^\prime - x_1) |x_1^\prime \rangle \,,
\ee 
where on the right hand side stands an ordinary delta function.

${\hat \psi}_T$ is defined on eigenstates $|x_0^\prime \rangle$ of ${\hat x}_0$:
\begin{gather}
{\hat x}_0 | x_0^\prime \rangle = x_0^\prime | {\hat x}_0 \rangle \,, \nonumber \\ 
{\hat \psi}_T ({\hat x_0}- x_0)| x_0^\prime \rangle = \psi_T (x_0^\prime - x_0) | x_0^\prime \rangle \,. 
\end{gather}
Here $\psi_T(\xi)$ is the characteristic function on the interval $\lbrack -\frac{T}{2} \,, \frac{T}{2} \rbrack$: 
\beqa
\psi_T(\xi) = 
\left \lbrace
\bea{ll}
1 & {\mathrm {for}}~~  |\xi| < \frac{T}{2} \\
0 & {\mathrm {for}}~~  |\xi| > 0 \,.
\ea
\right.
\eeqa

It is easy to show that ${\hat \gamma}$ is a positive operator and that $\omega_\gamma$ is a state.
Now $\psi_T^2 = \psi_T$. Hence for $\theta = 0$, we can write ${\hat \gamma}= \delta({\hat x_1} - x_1){\psi}_T({\hat x_0 
- x_0})$. The corresponding $\omega_{{\hat \gamma}}$ describes an experiment at spatial location $x_1$ which is
averaged uniformly over the time interval $\lbrack x_0- \frac{T}{2} \,, x_0 + \frac{T}{2} \rbrack$. For $\theta \neq 0$,
${\hat \gamma}$ is an approximation to such an experiment.

But for $\theta \neq 0$, ${\hat \gamma}$ does not have sharp spatial localization. If it did, then for 
\be
{\hat \alpha} = {\hat \delta}({\hat x}_1 - y_1) \,,
\ee
we should get
\be
\omega({\hat \alpha}) = \delta (x_1 - y_1) \,.
\ee

Instead we find (cf. Appendices 1 and 2)
\be
\omega_\gamma({\hat \alpha}) = 
\frac{2}{\pi} \frac{\theta}{T} \frac{1}{(x_1 - y_1)^2} \sin^2 \left 
[ \frac{T (x_1 -y_1)}{2 \theta} \right ]\,.
\label{eq:stateal}
\ee
As $\frac{T}{\theta} \rightarrow \infty$, it approaches to $\delta (x_1 - y_1)$ as it should.

It is important to emphasize that in order to interpret the action of
the state $\omega_{\hat \gamma}$ on $|{\hat \psi}|^2$ as the
measurement of intensity at a given spatial location, say $x_1$, it is
necessary to be able to localize $x_1$  with a reasonable precision
which becomes sharply localized as $\theta \rightarrow
0$. $\omega_{\hat \gamma}$ with density matrix ${\hat \gamma}$ does this
job perfectly: It approximately localizes the point where the
measurement is taken (c.f. equation (\ref{eq:stateal})) and
it has the correct commutative limit. 

A convenient way to calculate the traces such as those in (\ref{eq:stateal}) is to use the coherent states. Let 
\be 
a = \frac{{\hat x}_0 + i {\hat x}_1}{\sqrt{2 \theta}} \,, \quad a^\dagger 
= \frac{{\hat x}_0 - i {\hat x}_1}
{\sqrt{2 \theta}} \,, \quad \lbrack a \,, a^\dagger \rbrack = 1 \,,
\ee
and 
\be
|z \rangle = e^{\frac{1} {\sqrt{2 \theta}}(z a^\dagger -{\bar z} a)} | 0 \rangle \,,
\ee
where $a |z \rangle = \frac{z}{\sqrt{2 \theta}} |z \rangle$ and $\langle z | z  \rangle = 1$. We have
\be
\langle z| {\hat x}_\mu | z \rangle = x_\mu \,,
\ee
where $z = x_0 + i x_1$. 

We can now compute (\ref{eq:stateal}) using the resolution of identity
\be
{\bf 1} = \int \frac{d^2 z^\prime}{4 \pi \theta} | z^\prime \rangle \langle z^\prime | \,.
\ee
In particular we find that
\be
Tr {\hat \gamma} =  \int \frac{d^2 z^\prime}{4 \pi \theta} \langle z^\prime | {\hat \gamma} | z^\prime 
\rangle = \frac{T}{2 \pi \theta} \,.
\label{eq:trace1}
\ee   
Appendix 1 contains the details. 

Note that (\ref{eq:trace1}) diverges as $\theta\rightarrow 0$. That is because ${\hat \delta}({\hat x}_1 - x_1)$ is not
normalizable in the commutative limit just like a plane wave.

The object $I$ of our interest is the intensity as measured by $\omega_\gamma$:
\be
I = \omega_{\hat \gamma} (|{\hat \psi}|^2) \,.
\ee

We find, using coherent states for example, that
\beqa
I = \frac{Tr {\hat \gamma} |{\hat \psi}|^2}{Tr {\hat \gamma}} = 
\left \lbrace
\bea{cll}
2 \left[ 1 + \left ( 1 - \frac{ 2 \theta w }{T} \right ) \cos 2 w ( x_1 - \theta w) \right ] & \mbox{for} & 
2 \theta w  < T \,, \\
2 & \mbox{for} & 2 \theta w \geq T   
\ea
\right.
\label{eq:ilksonuc0}
\eeqa
as Appendix 3 shows.

This result is remarkable. It asserts that {\it there is no interference at all if $\frac{\theta w}{T} > \frac{1}{2}$!}
Thus the higher the frequency, the larger is the time of observation needed to perceive interference. There {\it is}
interference for $\frac{\theta w}{T} < \frac{1}{2}$, but its pattern is shifted and extrema modified depending on 
frequency and time of observation.  Figure 1 illustrates the phenomenon. We recover the usual pattern when 
$\frac{\theta w}{T} \rightarrow 0$, and in particular in the commutative limit. Variation of $I$ w.r.t. $2 \theta w$
is plotted in Figure 2.
\begin{figure}
\begin{center}
\includegraphics[width=0.5\textwidth, height=0.6\textheight, angle=-90]{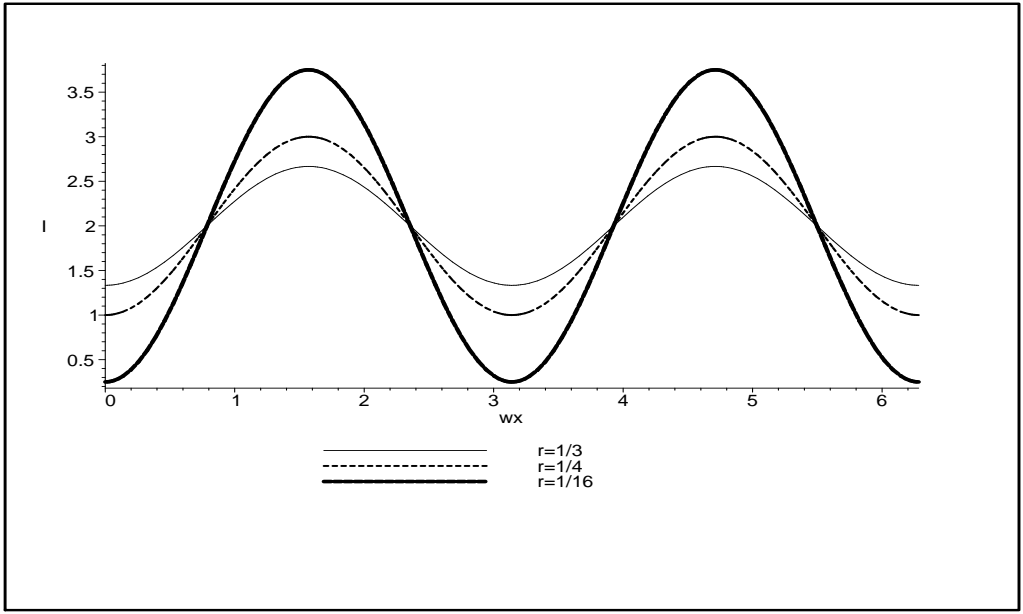}
\caption{Variation of intensity $I$ as a function of $w x_1$, for fixed $\frac{\theta w}{T} < \frac{1}{2}$. 
The plots are for $r= \frac{\theta w}{T} = \frac{1}{3} , \frac{1}{4} , \frac{1}{16}$ and  $\theta w^2 = \frac{\pi}{2}$. 
Clearly, as the ratio  $\frac{\theta w}{T}$ gets smaller, it converges to the commutative result.}
\end{center}
\end{figure}
\begin{figure}
\begin{center}
\includegraphics[width=0.5\textwidth, height=0.6\textheight, angle=-90]{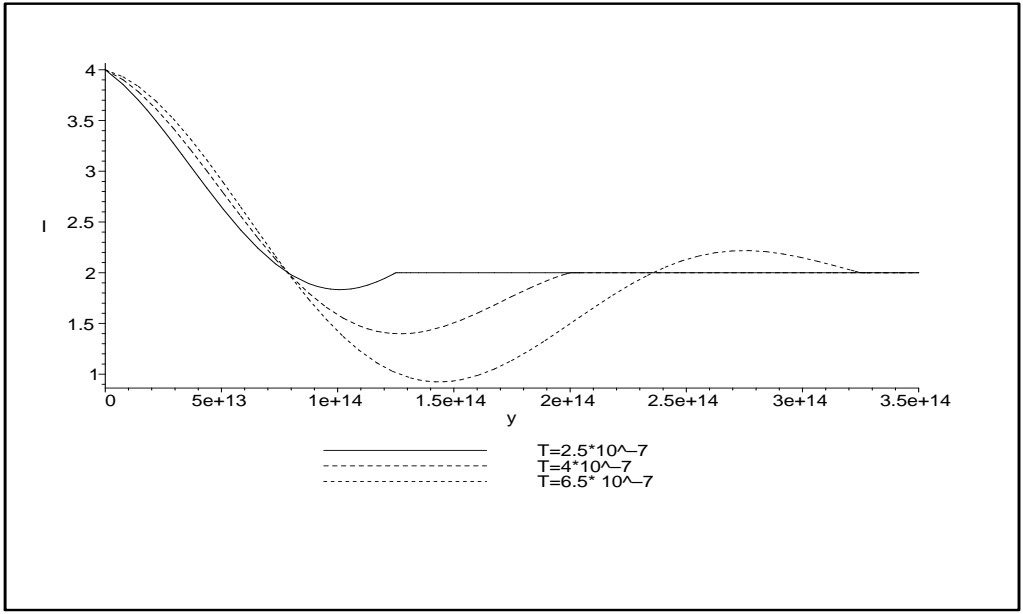}
\caption{Variation of intensity $I$ as a function of $y = \theta $. It
  is plotted at $x_1 = 0$, $\omega = 10^{-7} m^{-1}$ for the values of
$T = 2.5 \times 10^7, 4 \times 10^7, 6.5 \times 10^7 m $. 
Note that for $\theta =0$, $I=4$ and for any given $T$, $I$ takes the value $2$ and becomes
independent of $\theta$ at $\theta=\frac{T}{2} \times 10^7$ and thereon.}
\end{center}
\end{figure}

\vskip 0.3cm

{\it b) $d=3$}:

\vskip 0.2cm

We now consider the algebra ${\cal A}_\theta({\mathbb R}^3)$. A good problem to study is Young's double slit experiment.
For this, let us imagine a screen (a line) along the $2$-direction. The two slits are a distance $a$ apart and the line
joining them is also parallel to $2$-axis. This line is at a fixed distance from the screen. 

As spatial coordinates commute, there is no intrinsic difficulty in realizing an arrangement like the above with 
arbitrary precision. 

Let $\vec{n}= (1, 0)$. Then time-space non-commutativity is expressed as 
\be 
\lbrack {\hat x}_0, {\hat x}_1 \rbrack = i \theta  \,, \quad \lbrack {\hat x}_0, {\hat x}_2 \rbrack = 0 \,.
\ee

The wave is considered to be associated with a massless scalar field. Then as in (\ref{eq:planewave}), a plane wave is 
\be 
{\hat \psi}_1 = e^{i {\bf k} \cdot {\bf {\hat r}}} e^{- i k {\hat x}_0} \,, \quad {\bf {\hat r}} \equiv ( \hat{x}_1 \,, 
\hat{x}_2 ) \,,\quad {\bf k} \cdot {\bf {\hat r}} = k_i {\hat x}_i \,,\quad | {\bf k}| \equiv k = \sqrt{k_1^2 +k_2^2}\,.
\ee 

We can also spatially translate this wave by $\vec{a}$ and evolve it for time ${\tau}$ by applying $e^{i \vec{{\hat P}} 
\cdot \vec{a}} e^{-i {\hat P}_0 \tau}$, the result is still a plane wave:
\be
e^{-i {\hat P}_0 \tau}  e^{i \vec{{\hat P}} \cdot \vec{a}} \big ( e^{i {\bf k} \cdot {\bf {\hat r}}} e^{- i k {\hat x}_0}
\big)
= \big ( e^{i {\bf k} \cdot {\bf {\hat r}}} e^{- i k {\hat x}_0} \big) e^{i {\bf k} \cdot {\bf a}} e^{- i k \tau} \,.
\label{eq:phases1}
\ee

The last factor is the complex-valued phase of the wave. It lets us to unambiguously compare the phases of waves related 
by space-time translations. As the phase does not change if $\vec{a} = \frac{\vec{k}}{k} \tau$ for $k \neq 0$, the phase 
being zero for $k = 0$, we can say that the wave travels in the direction ${\vec k}$ as in the $\theta =0$ limit.

We want to consider the interference of the waves from the two slits at a point $P$ on the screen, assuming for
simplicity that they have the same frequency $w$.

Let ${\bf {\hat k}}$ and ${\bf {\hat k}}^\prime$ be the directions of propagation from the slits to $P$. Then the wave 
at $P$ is 
\begin{gather}
{\hat \psi} = {\hat \psi}_1 + {\hat \psi}_2  \,, \nonumber \\
{\hat \psi}_1 = e^{i {\bf k} \cdot {\bf {\hat r}}} e^{- i k {\hat x}_0} \,, \quad \quad {\hat \psi}_2 = e^{i {\bf k}^
\prime \cdot {\bf {\hat r}}^ \prime} e^{- i k {\hat x}_0} \,,
\end{gather}
where 
\be
{\bf k} = k {\bf {\hat k}} \,, \quad {\bf k}^\prime = k^\prime {\bf {\hat k}^\prime} \,, \quad {\bf {\hat r}}^\prime = 
{\bf {\hat r}} + {\bf {\hat a}} 
\ee
Here ${\bf a} = a (0,1)$ is the displacement of the primed slit relative to the other one. 

Note that the waves do not acquire phases as they arrive at $P$ from the slits a time $\tau$ later as the remark after
(\ref{eq:phases1}) shows.

We generalize the density matrix ${\hat \gamma}$ according to
\be
{\widehat \Gamma} = {\hat \psi}_T ({\hat x}_0 - x_0) {\hat \delta}({\hat x}_1 - x_1) {\hat {\hat \delta}} 
({\hat x}_2 - x_2) {\hat \psi}_T ({\hat x}_0 - x_0) \,, 
\label{eq:gendenmat}
\ee
where ${\hat \delta}({\hat x}_1 - x_1)$ and  ${\hat \psi}_T$ are given by (\ref{eq:deltaf}) and (\ref{eq:distpsi}) 
respectively, while ${\hat {\hat \delta}} ({\hat x}_2 - x_2)$ is a regularized version of the delta function 
centered at $x_2$(see below).
Consequently, the state generalizing (\ref{eq:densitymat}) is given by 
\be
\omega_{{\widehat \Gamma}}({\hat \alpha}) = \frac{Tr {\widehat \Gamma}{\hat \alpha}}{Tr {\widehat \Gamma}} \,.
\ee
Then the intensity $I$ due to ${\hat \psi}$ at $P$ is given by 
$\omega _{\widehat \Gamma}({\hat \psi}^\dagger {\hat \psi})$.

Note that the standard $\delta$-function ${\hat \delta} ( {\hat x}_2 - x_2)$
is not normalizable, having infinite trace. Hence, to regularize its 
contribution to the traces, we replace it, for example by 
${\hat {\hat \delta}}({\hat x_2} - x_2)=|\alpha \rangle \langle \alpha |$ 
where
\be
\langle x_2^\prime |\alpha \rangle \langle \alpha | x_2^\prime \rangle
\label{eq:ga1}
\ee
is a Gaussian of width $d$ centred at $x_2^\prime = x_2$. For this purpose we introduce
the momentum ${\hat P}_2$ conjugate to ${\hat x}_2$ with the eigenfunctions
$e^{i p_2^\prime (x_2^\prime - x_2)}$ and eigenvalues $p_2^\prime \in (-\infty \,, \infty)$. 
Consider
\beqa
\langle x_2^\prime |\alpha \rangle &=&  \frac{1}{\sqrt{2 \pi}} \frac{\sqrt{d}}{\pi^{1/4}} \int d p_2^\prime 
e^{i p_2^\prime (x_2^\prime - x_2)} e^{- \frac{1}{2} d^2 p_2^{\prime 2}} \nonumber \\
&=& \frac{1}{\pi^{1/4} \sqrt{d}} e^{- \frac{1}{2 d^2} (x_2^\prime- x_2)^2} \,.
\label{eq:ga2}
\eeqa
(\ref{eq:ga2}) will do the job: (\ref{eq:ga1}) is a Gaussian, which in the 
$d \rightarrow 0$ limit is a delta function centred at $x_2$, while $Tr {|\alpha \rangle \langle \alpha |} = 1$. 

With ${\hat {\hat \delta}} ({\hat x}_2 - x_2)= |\alpha \rangle \langle \alpha |$, traces 
can easily be computed in the basis $|z^\prime \rangle \otimes |x_2^\prime \rangle$ ($|z^\prime \rangle$ being the 
coherent states used in the previous section). We observe that $Tr {\widehat \Gamma} = \frac{T}{2 \pi \theta}$, while 
for the intensity $I$ we find
\begin{eqnarray}
&&I = \omega _{\widehat \Gamma}({\hat \psi}^\dagger {\hat \psi}) 
= \lim_{d \rightarrow 0} 2 \Big [ 1 + \Big ( 1 - \frac{\theta |k_1 - k_1^\prime|}{T} \Big) e^{-\frac{1}{4} 
d^2 (k_2- k_2^\prime)^2} \cos \Big ((k_1 -k_1^\prime) (x_1 - k \theta) \nonumber \\ 
&& \quad \quad \quad \quad \quad \quad \quad \quad \quad \quad \quad \quad \quad \hskip 4.5cm + (k_2- k_2^\prime) 
x_2 - k_2^\prime a \Big ) \Big ] \nonumber \\
&&= 2 \Big [ 1 + \Big ( 1 - \frac{\theta |k_1 - k_1^\prime|}{T} \Big) 
\cos \Big ((k_1 -k_1^\prime) (x_1 - k \theta) + (k_2- k_2^\prime) x_2 - k_2^\prime a \Big ) \Big ]
\quad \nonumber \\
&& \hskip 10cm \mbox{for} \quad \theta |k_1 - k_1^\prime| < T  \nonumber \,, 
\end{eqnarray}
\be
I = 2 \quad  \mbox{for} \quad \theta |k_1 - k_1^\prime| \geq T \,.
\label{eq:ilksonuc}
\ee
Note that the final result is independent of $d$.

This result is similar to the $d=$2 case. There is no interference at all for $ \theta |k_1 - k_1^\prime| \geq T$ and the
interference pattern is distorted for $\theta |k_1 - k_1^\prime| < T$ in a similar manner encountered in
(\ref{eq:ilksonuc0}).

When $\theta |k_1 - k_1^\prime| < T$, the dependence of $I$ on $(k_1 -k_1^\prime) x_1$ can be plotted for fixed values 
of $\frac{\theta |k_1-k_1^\prime|}{T}$. With suitable choice of the phases in (\ref{eq:ilksonuc}), the plots will look
similar to those in Figure 1.

\vskip 0.2cm  

{\it c) d =4 : Interference Phenomena from Cosmic Strings}

\vskip 0.2cm

As an example we here study interference of waves from a distant source caused by a cosmic string.
For simplicity we consider a straight cosmic string. The metric around
it is flat with deficit angle $8 \pi G \mu$, 
where $\mu$ is the mass per unit length of the string and $G$ is the gravitational constant.
\begin{figure}
\begin{center}
\includegraphics[width=0.6\textwidth, height= 0.25\textheight]{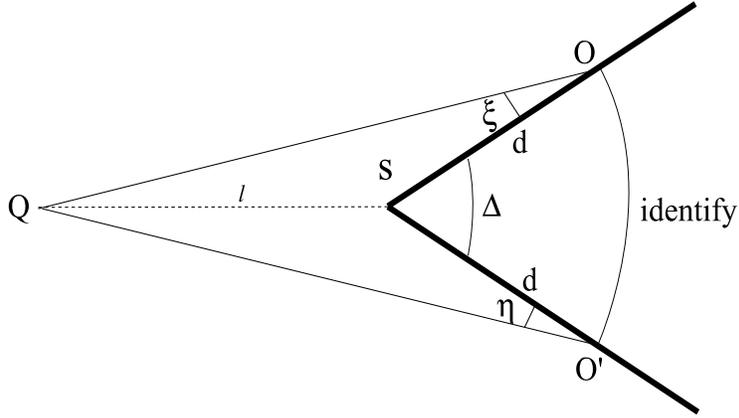}
\caption{Double image of $Q$ is observed by $O \equiv O^\prime$.}
\end{center}
\end{figure}

Figure 3 shows a spatial slice where $Q$ is the source and $S$ is the
string. We assume that $S$ is normal to the spatial slice as it is
sufficient for our purposes.
The string causes
deficit angles and requires the straight lines $SO$ and $SO^\prime$ to be identified. This identification causes a
double image of $Q$. $\Delta= \widehat{OSO^\prime}$ is the deficit
angle $ 8 \pi G \mu$. $Q$, $S$, $O$ and $O^\prime$
are on a plane ${\cal P}$. Calling $l$ the distance from the string to the source
and $d$ the distance from the observer to the string, the angular separation of the images is \cite{Vilenkin} 
\be
\Xi = \xi + \eta = \frac{l}{l +d} \Delta  \,,
\ee
(where $\xi$ and $\eta$ are explained by Figure 3.) 

We note that all the spatial coordinates commute with each other. Hence for the set-up above 
there is no intrinsic difficulty in defining the spatial directions and their orthogonality. 
They are exactly the same as in the commutative case.

It should be clear that the problem is effectively of two spatial
dimensions. On the spatial slice any fixed vector from the
string can be taken as the direction noncommuting with the time coordiante.   

Consider now the identified observers at $O$ and $O^\prime$. Call her $\varepsilon$. It is not hard to see from Figure 4 
that keeping her distance $d$ from the string $S$ fixed, she can observe double images due to $Q$
on all points on an arc $A$ of length 
$d \Delta $. (See Figure 4).
\begin{figure}
\begin{center}
\includegraphics[width=0.95\textwidth, height=0.35\textheight]{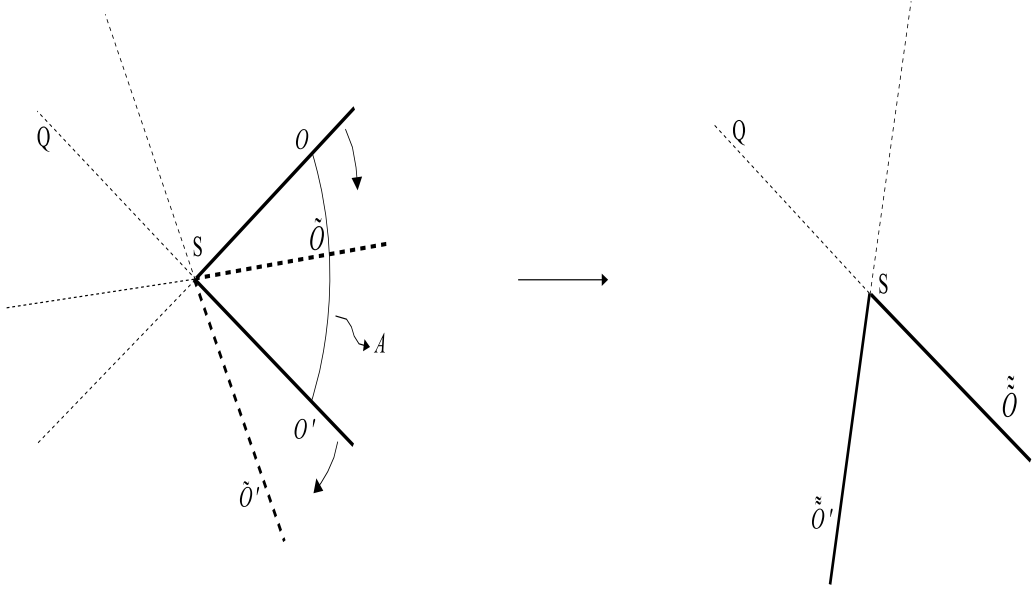}
\caption{The observer $\varepsilon$ at $O \equiv O^\prime$ can barely observe a double image of the source $Q$.
Keeping her distance from the string fixed, she can shift her position along the arc $A$ and can observe a double image
of $Q$ from $\tilde{O} \equiv \tilde{O}^\prime$. At $\tilde{\tilde{O}} \equiv \tilde{\tilde{O}}^\prime$ she again can
only barely observe a double image of $Q$.} 
\end{center}
\end{figure}
 
Suppose now that she can observe the intensity of waves of definite frequency $w$ coming from $Q$. Let us denote the wave
vectors of the two plane waves ${\hat \psi}_1$ and ${\hat \psi}_2$ emerging from $Q$ that reach the observer 
$\varepsilon$ at $O \equiv O^\prime$ by ${\bf k}$ and ${\bf k^\prime}$, respectively. 
At $O \equiv O^\prime$ the angle between 
${\bf k}$ and ${\bf k^\prime}$ is equal to $\Xi$. The intensity observed by $\varepsilon$ at $O \equiv 
O^\prime$ is then given by (\ref{eq:ilksonuc}), where $a$ in that formula (the separation between the slits) is now zero 
as there is only a single source $Q$. The observer $\varepsilon$ can shift the observed intensity by shifting her 
position on the arc $A$. Thus we can think of $A$ as a screen, where the interference pattern is recorded. 

After some trigonometry we find that (see Appendix 4)
\be
|k_1 - k_1^\prime| = \Big | (k_2 + k_2^\prime) \tan \frac {\Xi}{2} \Big | \,.
\label{eq:trigo}
\ee  

From this result it is possible to estimate an upper bound on $\frac{\theta}{T}$ in order that the two waves interfere. 
Recall that for this we must have $\frac{\theta |k_1 - k_1^\prime|} {T} < 1$. Substituting for 
$|k_1 - k_1^\prime|$ from (\ref{eq:trigo}), we get $\frac{\theta |k_2+ k_2^\prime|}{T} < \big |\cot \frac{\Xi}{2}
\big|$. It is known that double images of quasars are usually separated by a few arc seconds \cite{castles}.
Taking for example an angular separation of 5 arc seconds puts the bound $\frac{\theta |k_2+ k_2^\prime|}{T} 
< 8.25 \times 10^4$. We note that $|k_2+ k_2^\prime| = \alpha k$ where $0 \leq \alpha < 2$. Therefore, 
$\frac{\theta }{T} < \frac{8.25 \times 10^4}{\alpha k}$ and for a given $k$ the lowest upper bound will be approached as 
$\alpha \rightarrow 2$. Finally, suppose that a wavelength at the red end of the visible spectrum is observed, say at
$\lambda= \frac{2 \pi}{k} =700 nm$. Then we find $\frac{\theta }{T} < 4.6 \times 10^{-3} m$. 
Only under this condition is the interference observable for light with
wavelength $\lambda =700 nm$.

\section{Conclusions}

In this work we have studied the general theory of waves in Groenewold-Moyal spacetimes where time and space coordinates do 
not commute. We have given the rules for the measurement of their intensity and applied them to interference 
and diffraction phenomena in spacetimes of dimensions $d \leq 4$. The latter produced novel physical results.
Namely, we found out that for observation times $T$ which are so brief that $T \leq 2 \theta w$, no interference can be observed. For larger times, the interference pattern is deformed and depends on $\frac{\theta w}{T}$. It approaches the commutative pattern only when $\frac{\theta w}{T} \rightarrow 0$. These results are given concretely by the equations (\ref{eq:ilksonuc0}) and (\ref{eq:ilksonuc}) for $d=2$ and $d=3$, respectively. Finally, we have used these results to discuss the interference of stellar light due to cosmic strings, where with the help of the present stellar data we have
estimated that for a given $k$, we must have $\frac{\theta }{T}
\lesssim \frac{4 \times 10^4}{k}$ to observe interference.

\vskip 2em
  
{\bf Acknowledgments}

\vskip 2em

A.P.B and S.K. would like to thank A. Pinzul and B. Qureshi for discussions. 
The work of A.P.B and S.K. are supported in part by the DOE grant DE-FG02-85ER40231 and the NSF under contract number
INT9908763. S.K. acknowledges support from the IRCSET postdoctoral fellowship.
A part of this work was done during K.S.G's visit to the Abdus Salam ICTP,
Trieste, Italy within the framework of the Associateship Programme of the 
Abdus Salam ICTP and KSG would like to thank the Associateship
Scheme of the Abdus Salam ICTP for support during his visit.

\vskip 2em
\appendix{{\bf Appendix 1}}
\renewcommand{\theequation}{A.\arabic{equation}}
\setcounter{equation}{0}
\renewcommand{\thefigure}{A. \arabic{figure}}
\setcounter{figure}{0}
\vskip 2em

In this Appendix we calculate $Tr {\hat \gamma}$ where ${\hat \gamma}$ is
given by (26). In the coherent state basis $| z^\prime \rangle $  we have
\beqa
Tr {\hat \gamma} &=&
\int \frac{d^2 z^\prime}{4 \pi \theta}
\langle z^\prime | {\hat \gamma} | z^\prime \rangle \nonumber \\
&=& \int \frac{d^2 z^\prime}{4 \pi \theta}
\langle z^\prime | {\hat \psi}_T({\hat x_0}- x_0)
{\hat \delta} ( {\hat x}_1 - x_1)
{\hat \psi}_T({\hat x_0}- x_0) | z^\prime \rangle.
\label{eq:A1} 
\eeqa
Using (27) and (28) in (\ref{eq:A1}), we get 
\beqa
Tr {\hat \gamma} &=& \int d \lambda_1 d
\lambda_2~
Tr ~{\hat \delta}({\hat x}_0 - x_0 - \lambda_1)~ {\hat \delta} ( {\hat x}_1 -
x_1) {\hat \delta}({\hat x}_0 - x_0 - \lambda_2)  \nonumber \\
&=& \frac{1}{({2 \pi})^3}\int  d \lambda_1 d
\lambda_2 d k_1 d k_2 d k_3~
Tr ~e^{i k_1 ({\hat x}_0 - x_0 - \lambda_1)} e^{i k_2 ({\hat x_1}- x_1)}
e^{i k_3 ({\hat x}_0 - x_0 - \lambda_2)}  \\
&=& \frac{1}{({2 \pi})^3}\int  d \lambda_1 d\lambda_2 d k_1
d k_2 d k_3
e^{i[\theta k_2 k_3 - k_1(x_0 + \lambda_1) - k_2 x_1 - k_3(x_0 + \lambda_2)]}
Tr  e^{i (k_1 + k_3) {\hat x_0}} e^{i k_2  {\hat x_1}} \nonumber
\label{eq:A2} 
\eeqa
Using (36) and in the coherent state basis we have 
\beqa
Tr  e^{i (k_1 + k_3) {\hat x_0}} e^{i k_2  {\hat x_1}}
&=& \langle z^{\prime} | e^{i (k_1 + k_3) {\hat x_0}} e^{i k_2  {\hat x_1}} 
|z^{\prime} \rangle \nonumber \\
&=&  \int \frac{dx_0^{\prime} dx_1^{\prime}}{2 \pi
\theta}
e^ {[ \frac{\theta}{4} \{ (k_1 + k_3)^2 + k_2^2 \} 
- i \frac{\theta}{2} (k_1 + k_3)k_2 ]} e^{i (k_1 + k_3) x^{\prime}_0}
e^{i k_2 x^{\prime}_1} \nonumber \\
&=& \frac{2 \pi}{\theta} e^ {[ \frac{\theta}{4} \{ (k_1 + k_3)^2 + k_2^2
\} - i \frac{\theta}{2} (k_1 + k_3)k_2 ]} \delta (k_1 + k_3)~ \delta(k_2)
\nonumber \\
&=& \frac{2 \pi}{\theta} \delta (k_1 + k_3)~ \delta(k_2),
\label{eq:A3} 
\eeqa
where we have used $z^{\prime} = x_0^{\prime} + i x_1^{\prime}$ and
$d^2 z^{\prime} = 2 dx_0^{\prime}dx_1^{\prime}$.
Substituting (\ref{eq:A3})in (A.2) and remembering that $ - \frac{T}{2} \leq
\lambda_i \leq \frac{T}{2}, ~~i = 1,2$, we get
\beqa
Tr {\hat \gamma}
&=& \frac{1}{ 4 \pi^2 \theta} \int d\lambda_1 d \lambda_2 d k_3
e^{i k_3 (\lambda_1 + \lambda_2)} \nonumber \\
&=& \frac{1}{2 \pi \theta} \int d\lambda_1 d \lambda_2 \delta (\lambda_1 +
\lambda_2) \nonumber \\
&=& \frac{T}{2 \pi \theta}
\label{eq:A4}
\eeqa

\vskip 2em

\appendix{{\bf Appendix 2}}

\vskip 2em
In this Appendix we give the derivation of (35). Using (26), (27), (28) and (33) we get
\beqa
Tr {\hat \gamma} {\hat \alpha} &=&
Tr {\hat \psi}_T({\hat x_0}- x_0) {\hat \delta} ( {\hat x}_1 - x_1) 
{\hat \psi}_T({\hat x_0}- x_0) 
{\hat \delta} ( {\hat x}_1 - y_1) \nonumber \\
&=& \frac{1}{(2 \pi)^4} \int d \mu ~ Tr 
e^{i k_1 ({\hat x_0}- x_0 - \lambda_1)} e^{i k_2 ( {\hat x}_1 - x_1)}
e^{i k_1 ({\hat x_0}- x_0 - \lambda_2)} e^{i k_2 ( {\hat x}_1 - y_1)} \\
&=& \frac{1}{(2 \pi)^4} \int d \mu ~
e^{i [\theta k_2 k_3 - k_1 (x_0 +  \lambda_1) - k_2 x_1 - k_3  (x_0 +
\lambda_2) - k_4 y]} ~ Tr e^{i (k_1 + k_3) {\hat x_0}}
e^{i (k_2 + k_4) {\hat x_1}}, \nonumber 
\label{eq:A5}
\eeqa
where $d \mu = d \lambda_1 d \lambda_2 dk_1 dk_2 dk_3 dk_4$ and 
$ - \frac{T}{2} \leq \lambda_i \leq \frac{T}{2}, ~~i = 1,2$.
Using (36) and in the coherent state basis we get
\beqa
Tr e^{i (k_1 + k_3) {\hat x_0}} e^{i (k_2 + k_4) {\hat x_1}} &=&
\frac{2 \pi}{\theta} e^{\frac{\theta}{4}[(k_1 + k_3)^2 + (k_2 + k_4)^2]}
\delta (k_1 + k_3) \delta (k_2 + k_4) \nonumber \\
&=& \frac{2 \pi}{\theta} \delta (k_1 + k_3) \delta (k_2 + k_4)
\label{eq:A6}
\eeqa
where we have used a  technique similar to that was used to derive (\ref{eq:A3}). Using
(\ref{eq:A6}) in (A.5) we get
\beqa
Tr {\hat \gamma} {\hat \alpha} &=&
\frac{1}{(2 \pi \theta)^2} \int d \lambda_1 d \lambda_2 ~
e^{i \lambda_1 \frac{(y_1 - x_1)}{\theta}}
e^{i \lambda_2 \frac{(y_1 - x_1)}{\theta}} \nonumber \\
&=& \frac{1}{\pi^2 (x_1 - y_1)^2}~ \sin^2 \left [ \frac{T (x_1 - y_1)}{2
\theta} \right ]
\eeqa
Using (61) and (58) in (25) we get 
\be
\omega_\gamma({\hat \alpha}) =
\frac{2}{\pi} \frac{\theta}{T} \frac{1}{(x_1 - y_1)^2} \sin^2 \left
[ \frac{T (x_1 -y_1)}{2 \theta} \right ].
\ee

\vskip 2em

\appendix{{\bf Appendix 3}}

\vskip 2em

In this Appendix we provide the derivation of the intensity formula (42).
Using (23), we get
\beqa
Tr {\hat \gamma} |{\hat \psi}|^2 &=&
2 Tr {\hat \gamma} + e^{-2ik^2 \theta} Tr {\hat \gamma} e^{2ik {\hat x_1}} +
e^{2ik^2 \theta} Tr {\hat \gamma} e^{- 2ik {\hat x_1}} \nonumber \\
&=& 2 \left [ Tr {\hat \gamma} + {\rm Re} ( e^{-2ik^2 \theta} Tr {\hat
\gamma} e^{2ik {\hat x_1}} ) \right ] .
\label{eq:A9}
\eeqa
Substituting for ${\hat \gamma}$ from (26), we can write
\be
Tr {\hat \gamma} e^{2ik {\hat x_1}} =
\frac{1}{(2 \pi)^3} \int d \mu 
e^{-i [ k_2 k_3 \theta + k_1 (x_0 + \lambda_1) + k_2 x_1 
+ k_3 (x_0 + \lambda_2)]}~ Tr e^{i (k_1 + k_3) {\hat x_0}}
e^{i (k_2 + 2k) {\hat x_1}},
\ee
where $ d \mu = d \lambda_1  d \lambda_2 dk_1 dk_2 dk_3$. Using (36) and in
the coherent state basis we get
\beqa
Tr e^{i (k_1 + k_3) {\hat x_0}} e^{i (k_2 + 2k) {\hat x_1}} &=& 
\frac{2 \pi}{\theta} e^{\frac{\theta}{4}[(k_1 + k_3)^2 - (k_2 + 2k)^2]}
\delta (k_1 + k_3) \delta (k_2 + 2 k) \nonumber \\
&=& \frac{2 \pi}{\theta} \delta (k_1 + k_3) \delta (k_2 + 2 k) .
\eeqa
Hence
\beqa
e^{-2ik^2 \theta} Tr {\hat \gamma} e^{2ik {\hat x_1}} &=&
\frac{e^{2i (k x_1 - k^2 \theta)}}{(2 \pi ) \theta}  
\int_{-\frac{T}{2}}^{ \frac{T}{2}} d \lambda_2
\int_{-\frac{T}{2}}^{ \frac{T}{2}} d \lambda_1 
~\delta (\lambda_1 - \lambda_2 - 2 \theta k) \nonumber \\
&=& 
\frac{e^{2i (k x_1 - k^2 \theta)}}{(2 \pi ) \theta}
\int_{-\frac{T}{2}}^{\frac{T}{2} - 2 k \theta} d \lambda_2 .
\label{eq:A12}
\eeqa

Remembering that $\theta k > 0$, we see that the integrand in (\ref{eq:A12}) 
vanishes if $2 \theta k > T$. If $2 \theta k < T$ instead, we have
\beqa
e^{-2ik^2 \theta} Tr {\hat \gamma} e^{2ik {\hat x_1}} &=&
\frac{e^{2i (k x_1 - k^2 \theta)}}{(2 \pi ) \theta}
\int_{-\frac{T}{2}}^{\frac{T}{2} - 2 k \theta} d \lambda_2 \nonumber \\
&=& 
\frac{e^{2i (k x_1 - k^2 \theta)}}{ \pi \theta} \left ( \frac{T}{2} - 
\theta k \right ) .
\label{eq:A13}
\eeqa 
From (\ref{eq:A9}), (\ref{eq:A12}), (\ref{eq:A13}) and (\ref{eq:A4}) we finally get
\beqa 
I = \frac{Tr {\hat \gamma} |{\hat \psi}|^2}{Tr {\hat \gamma}} =  
\left \lbrace                                                   
\bea{cll} 
2 \left[ 1 + \left ( 1 - \frac{ 2 \theta w }{T} \right ) 
\cos 2w( x_1 - \theta w) \right ] & \mbox{for} &  
2 \theta w  < T \,, \\ 
2 & \mbox{for} & 2 \theta w \geq T   
\ea 
\right. ,
\eeqa 
where we have used $w = k$.

\vskip 2em

\appendix{{\bf Appendix 4}}
\vskip 2em

In this appendix we give a derivation of the result (\ref {eq:trigo}).
\begin{figure}
\begin{center}
\includegraphics[width=0.8\textwidth, height=0.25\textheight]{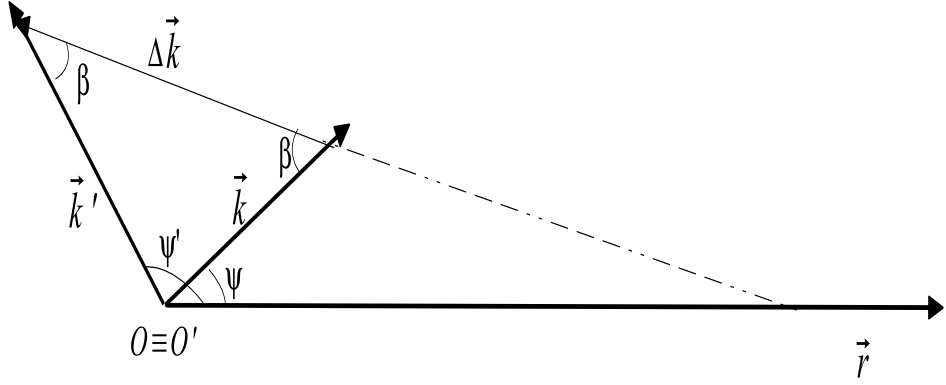}
\caption{The diagram provides the definition of angles $\psi \,, \psi^\prime$ and $\beta$.}
\end{center}
\end{figure}
From the figure above we see the following relations between the angles
\begin{gather}
\psi^\prime - \psi = \Xi \,, \\
2 \beta + \Xi = 180 \quad \Rightarrow \quad  \beta = 90 - \frac{\Xi}{2} \,, \\
2 \beta + \psi^\prime - \psi = 180 \quad \Rightarrow   \quad \psi^\prime + \beta = 180 - \beta + \psi \,.  
\label{eq:angles}
\end{gather}

Noting that $\cos \psi = \frac{{\bf k} \cdot {\bf r}}{k r} \,, 
\sin \psi = \frac{ {\hat {n}}.({\bf k} \times {\bf r})}{k r}$ (${\hat {n}}$ being
the unit normal in the direction of ${\bf k} \times {\bf r}$)
and $|k| = |k^\prime| = k$, we explicitly have
\be
\cos \psi = \frac{1}{k}  \left( k_1 \frac{x_1}{r} + k_2 \frac{x_2}{r} \right) \, \quad
\sin \psi = \frac{1}{k}  \left( k_1 \frac{x_2}{r} - k_2 \frac{x_1}{r} \right) \,,
\ee
and  similarly
\be
\cos \psi^\prime = \frac{1}{k}  \left( k_1^\prime \frac{x_1}{r} + k_2^\prime \frac{x_2}{r} \right) \, \quad
\sin \psi^\prime = \frac{1}{k}  \left( k_1^\prime \frac{x_2}{r} - k_2^\prime \frac{x_1}{r} \right) \,.
\ee
Using the above relations we find that
\begin{gather}
\cos ( \beta + \psi^\prime) = \frac{1}{k} \left \lbrack  \left ( k_1^\prime \sin \frac{\Xi}{2} + k_2^\prime \cos 
\frac{\Xi}{2} \right) \frac{x_1}{r} +  \left( k_2^\prime \sin \frac{\Xi}{2}  - k_1^\prime \cos \frac{\Xi}{2} \right) 
\frac{x_2}{r} \right \rbrack \,, \nonumber \\
\cos ( 180 - \beta + \psi ) = \frac{1}{k} \left \lbrack  \left ( - k_1 \sin \frac{\Xi}{2} + k_2 \cos \frac{\Xi}{2}
\right) \frac{x_1}{r} +  \left(- k_2 \sin \frac{\Xi}{2}  - k_1 \cos \frac{\Xi}{2} \right) \frac{x_2}{r} \right \rbrack 
\,.
\end{gather}
Due to (\ref{eq:angles}) the difference of the the cosines above vanish. For this to happen, we observe 
that the coefficients of $\frac{x_1}{r}$ and $\frac{x_2}{r}$ must separately vanish. Thus the coefficients of  
$\frac{x_2}{r}$ must satisfy
\be
(k_2 + k_2^\prime) \sin \frac{\Xi}{2} = (k_1^\prime - k_1) \cos \frac{\Xi}{2} \,,
\ee
and hence (\ref{eq:trigo}).

\end{document}